\documentclass[epjc,twocolumn]{revtex4}%
\usepackage{amsfonts}
\usepackage{amsmath}
\usepackage{amssymb}
\usepackage{graphicx}%
\begin{document}
\title{Shielding of Penrose superradiance in optical black holes}
\author{Hongbin Zhang}
\author{Baocheng Zhang}
\email{zhangbaocheng@cug.edu.cn}
\affiliation{School of Mathematics and Physics, China University of Geosciences, Wuhan
430074, China}
\keywords{radiation shielding, Penrose superradiance, optical black holes}
\begin{abstract}
We investigate the effect of superradiance shielding for the analogue rotating
black holes simulated by optical vortices by calculating the radial motion of
massless particles in such spacetime background. We add the conditions
$E<L\Omega_{r_{e}}$ and $L>0$ to judge the classically forbidden region of
superradiance. It is found that the superradiance forbidden region exists near
the static limit inside the ergosphere, which will limit the classical Penrose
process for the particles with some specific energies and angular momenta.
Once these particles satisfying the superradiance conditions are measured at
the outside of the ergosphere, this shows that the Penrose process can be quantum.

\end{abstract}
\maketitle

\section{Introduction}

The superradiance of rotating black holes was proposed \cite{penrose1971}
firstly by Penrose in the year of 1971. This can be used to extract the
rotational energy of a rotating black hole in such a way that an object is
emitted into the ergosphere where it is split into two pieces, in which one
has negative energy and the other one can escape from the ergosphere with a
positive energy gain. Furthermore, Zel'dovich \cite{zel1971} gave the
conditions for the existence of rotational superradiance,
\begin{equation}
\omega<n\Omega. \label{rsc}%
\end{equation}
where $\Omega$ is the angular velocity of the rotating black holes, $\omega$
is the angular frequency of the incident wave, and $n$ is the wave winding
number concerning the rotation axis. The superradiance can occur for the
classical or quantum waves. However, radiation screening \cite{Prodanov2014}
for the particles with certain energies and angular momenta exists inside and
outside the ergosphere for the Kerr spacetime \cite{mc2016hawking} and
Kerr-Newman spacetime \cite{gillani2021}, which restricts the motion of the
classical particles in the ergosphere and so restricts the occurrence of
classical Penrose process. For these restricted particles, the Penrose
superradiance can occur by the quantum tunneling. These particles can be
Hawking radiations from the event horizon, and can also be that emitted into
the ergosphere from the outside \cite{mc2016hawking}.

In the past several years, the Penrose superradiance was studied in the
analogue rotating black holes \cite{basak2003,Basak2003,prain2019,sp21,pgw22}.
The concept of the analogue black holes was put forward initially in 1981
\cite{unruh1981}, based on the relation between the motion of sound waves in a
convergent fluid flow and the motion of a scalar field in the background of
Schwarzschild spacetime, and the analog horizon is defined by equating the
velocity of the fluid with the local sound velocity in this fluid. The
analogue Hawking radiation can be emitted from the analogue horizon, which has
been studied in many different physical systems
\cite{weinfurtner2011,vieira2017,kolobov2021,munoz2019,fabbri2021,philbin2008,robertson2012}%
. Besides simulating the Schwarzschild black holes, rotating black holes were
also simulated using the physical systems \cite{blv11}, and some related
phenomena such as superradiance \cite{prain2019}, black-hole bombs
\cite{berti2004}, and scalar clouds \cite{ciszak2021} were investigated. In
this paper, we focus on the superradiance of analogue rotating black holes,
which could be measured experimentally
\cite{torres2017,braidotti2020,braidotti2022} in some physical systems. It is
pointed out that there will be no superradiance when the angular momentum is
negative in the optical systems \cite{braidotti2020,braidotti2022}. In this
paper, we aim to explore whether the radiation shielding regions exist under
the background of analogue rotating black holes simulated by the optical
vortices and their influence on the Penrose superradiance.

The paper is organized as follows. In Sec. II, we review how the metric of the
analogue black holes is derived from the nonlinear Schr\"{o}dinger equation
and investigate the radiation shielding regions for the analogue rotating
black holes. In Sec. III, we use the conditions for the occurrence of
superradiance to discuss the superradiance forbidden regions in the optical
rotating black holes. Then, we study the turning (boundary) points of the
classically forbidden region using an experimentally generated analogue
rotating spacetime with a static limit but without an event horizon in Sec.
IV. We also calculate the probability of the classically forbidden particle
tunneling through the ergosphere. Finally, we give the conclusion in Sec. V.

\section{Radiation shielding}

In order to investigate the radiation shielding phenomenon under the
background of analogue gravity, we first introduce how the metric of the
analogue black holes is obtained in the physical system of optical vortices.
Start with the nonlinear Schr\"{o}dinger equation (NLSE) \cite{rwb02} in the
paraxial approximation, which governs the evolution of the electric field
$\epsilon(x,y,z)$ of the vortex beam as,
\begin{equation}
\partial_{z}\epsilon=\frac{i}{2k}\nabla_{\perp}^{2}\epsilon-i\frac{kn_{2}%
}{n_{0}}\epsilon|\epsilon|^{2},\label{eq:NLSE}%
\end{equation}
where $z$ is the propagation direction, $k=\left(  2\pi n_{0}\right)
/\lambda$ is wave number along the $z$-direction, $n_{0}$ is the linear
refractive index, and $n_{2}$ is the nonlinearity coefficient. In this
equation, $z=c\tilde{t}/n_{0}$ is equivalent to time due to the constant light
speed. The first term on the right-hand side describes the diffraction effect,
and the second term describes the self-defocusing effects. If the electric
field is expressed as $\epsilon=\sqrt{\rho_{0}}e^{i\phi}$, the NLSE becomes
the continuity and Euler equations,
\begin{align}
\partial_{\tilde{t}}\rho+\nabla\cdot(\rho\mathbf{v}) &  =0,\\
\partial_{\tilde{t}}\psi+\frac{1}{2}v^{2}+\frac{c^{2}n_{2}}{n_{0}^{3}}%
\rho-\frac{c^{2}}{2k^{2}n_{0}^{2}}\frac{\nabla^{2}\rho^{1/2}}{\rho^{1/2}} &
=0,
\end{align}
where $c$ is the speed of light, the optical intensity $\rho$\ corresponds to
the fluid density, ${v}=\frac{c}{kn_{0}}\nabla\phi\equiv\nabla\psi$ is the
fluid velocity, and the fourth term is the quantum pressure which is usually
ignored in the linearized process for the derivation of the analogue metric.
Linearizing these equations with $\rho=\rho_{0}+\epsilon\rho_{1}$ and
$\psi=\psi_{0}+\epsilon\psi_{1}$, it is obtained that $(\dfrac{\rho_{0}}%
{c_{s}})^{2}(-\partial_{t}^{2}\psi_{1}-\partial_{t}\delta_{ij}\nu_{j}%
\partial_{j}\psi_{1}+c_{s}^{2}\partial_{i}\delta_{ij}\partial_{j}\psi
_{1}-\partial_{i}\delta_{ij}\nu_{j}\partial_{t}\psi_{1}-\partial_{i}%
v_{i}\partial_{j}v_{j}\psi_{1})=0$ where $c_{s}^{2}=c^{2}n_{2}\rho_{0}%
/n_{0}^{3}$ is the local speed of sound and $i,j=1,2$. Rewrite the equation
with the form, $\nabla^{2}\psi_{1}=\left(  1/\sqrt{-g}\right)  \partial_{\mu
}\left(  \sqrt{-g}g^{\mu\nu}\partial_{\nu}\psi_{1}\right)  $, one can obtain
the metric as \cite{marino2008},
\begin{equation}
ds^{2}=(\frac{\rho_{0}}{c_{s}})^{2}[-(c_{s}^{2}-v_{t}^{2})d\tilde{t}%
^{2}-2v_{r}drd\tilde{t}-2v_{\theta}rd\tilde{\theta}d\tilde{t}+dr^{2}%
+(rd\tilde{\theta})^{2}],\label{abm}%
\end{equation}
where $v_{r}=\partial_{r}\psi_{0}$, $v_{\theta}=\frac{1}{r}\partial
_{\tilde{\theta}}\psi_{0}$ are the radial and tangential velocity components,
and $v_{t}^{2}=v_{r}^{2}+v_{\theta}^{2}$ is the total velocity.

Making the time and angle transformations, $dt=d\tilde{t}+\frac{|v_{r}%
|}{(c_{s}^{2}-v_{r}^{2})}dr$, and $d\theta=d\tilde{\theta}+\frac
{|v_{r}|v_{\theta}}{r(c_{s}^{2}-v_{r}^{2})}dr$, the analogue metric
(\ref{abm}) becomes
\begin{equation}
ds^{2}=(\frac{\rho_{0}}{c_{s}})^{2}[-(c_{s}^{2}-v_{t}^{2})dt^{2}+\frac
{c_{s}^{2}}{c_{s}^{2}-v_{r}^{2}}dr^{2}+(rd\theta)^{2}-2v_{\theta}rd\theta dt].
\label{akm}%
\end{equation}
It is similar to the Kerr metric in general relativity, and has the event
horizon when $c_{s}=v_{r}$ and the static limit when $c_{s}=v_{t}$.

Such analogue metric has been realized by the optical vortex as given in Ref.
\cite{vmf18} and the superradiance was studied in Ref. \cite{marino2009}. In
this paper, we take the same parameters as in Ref. \cite{marino2009} to
analyze the screening effect of the analogue spacetime. Take the electric
field $\epsilon=\sqrt{\rho_{0}}exp(im\theta-2i\pi\sqrt{\frac{r}{r_{0}}})$
where $\rho_{0}$ is a constant optical intensity, and $r_{0}=100\mu m$ is an
experimental parameter to form the optical black hole. $v_{r}=-\frac{c\pi
}{kn_{0}\sqrt{r_{0}r}}$, $v_{\theta}=\frac{cm}{kn_{0}r}$, and $c_{s}%
=\sqrt{\frac{c^{2}n_{2}\rho_{0}}{n_{0}^{3}}}$ are the radial, angular, and
sound velocities, respectively. Moreover, the refractive index change takes
$n_{2}\rho_{0}=2\times10^{-6}$ which gives the sound velocity as $c_{s}%
=\sqrt{18}\times10^{5}m/s$. Other parameters take $\xi=\frac{c\pi}{kn_{0}%
c_{s}}=275\mu m$, $v_{\theta}=\frac{m\xi c_{s}}{\pi r}$, and $v_{r}=-\frac{\xi
c_{s}}{\sqrt{r_{0}r}}$. Thus, the event horizon of the optical black hole
locates at $r_{e}\approx756\mu m$, and the static limit at $r_{s}\approx893\mu
m$, with the topological charge $m=4$ of the vortex beam. With these
expressions in mind, the contravariant form of the analogue metric (\ref{akm})
can be written as%
\begin{equation}
\mathbf{g^{\mu\nu}}=\left(
\begin{array}
[c]{ccc}%
\frac{1}{\frac{\xi^{2}}{r_{0}r}-1} & 0 & \frac{\frac{m\xi}{\pi r}}{r\left(
\frac{\xi^{2}}{r_{0}r}-1\right)  }\\
0 & 1-\frac{\xi^{2}}{r_{0}r} & 0\\
\frac{\frac{m\xi}{\pi r}}{r\left(  \frac{\xi^{2}}{r_{0}r}-1\right)  } & 0 &
\frac{-1+\frac{\xi^{2}}{r_{0}r}+\left(  \frac{m\xi}{\pi r}\right)  ^{2}}%
{r^{2}\left(  \frac{\xi^{2}}{r_{0}r}-1\right)  }%
\end{array}
\right)  . \label{mte}%
\end{equation}

In the following, we will calculate the equations of motion for particles
moving in the analogue spacetime following the method by S. Chandrasekhar for
the geodesic motion in the spacetime of Kerr black holes \cite{chandr1983}.
Considering a particle with the action $S$ moving freely in the spacetime
described with the metric (\ref{akm}), the Hamilton-Jacobi (HJ) equation is
given as
\begin{equation}
2\frac{\partial S}{\partial\tau}=g^{\mu\nu}\frac{\partial S}{\partial x^{\mu}%
}\frac{\partial S}{\partial x^{\nu}},
\end{equation}
where $\tau$ is the proper time of the particle. Assuming the energy and the
angular momentum of the particle are constant $E$ and $L$, the action can be
expressed as
\begin{equation}
S=-\frac{1}{2}\delta\tau-Et+L\theta+S_{r}(r), \label{aep}%
\end{equation}
where $\delta$ is a constant and determines whether the motion equation of the
particle describes a time-like geodesic ($\delta>0$) or a light-like geodesic
($\delta=0$). Substituting metric tensor $\mathbf{g^{\mu\nu}}$ (\ref{mte}) and
the action (\ref{aep}) into the HJ equation, we have
\begin{align}
&  r^{2}\left(  \frac{\xi^{2}}{r_{0}r}-1\right)  ^{2}\left(  \frac{\partial
S_{r}(r)}{\partial r}\right)  ^{2}=r^{2}E^{2}+r^{2}\left(  \frac{\xi^{2}%
}{r_{0}r}-1\right)  \delta\nonumber\\
&  +\left(  -1+\frac{\xi^{2}}{r_{0}r}+\left(  \frac{m\xi}{\pi r}\right)
^{2}\right)  L^{2}-\frac{2m\xi EL}{\pi}. \label{hje}%
\end{align}

Integrating the Eq. (\ref{hje}), we obtain
\begin{equation}
S_{r}(r)=\int^{r}\frac{\sqrt{V_{eff}}}{r\left(  \frac{\xi^{2}}{r_{0}%
r}-1\right)  }dr,
\end{equation}
where the effective potential is expressed with the form
\begin{align}
V_{eff}  &  =r^{2}E^{2}+r^{2}\left(  \frac{\xi^{2}}{r_{0}r}-1\right)
\delta\nonumber\\
&  +\left(  -1+\frac{\xi^{2}}{r_{0}r}+\left(  \frac{m\xi}{\pi r}\right)
^{2}\right)  L^{2}-\frac{2m\xi EL}{\pi}. \label{epe}%
\end{align}

With the expression of $S_{r}(r)$ in the action (\ref{aep}), we can obtain the
equations of motion of the particle as%
\begin{equation}
r\frac{dr}{d\tau}=\sqrt{V_{eff}},
\end{equation}%
\begin{equation}
r\frac{dt}{d\tau}=\frac{1}{r\left(  \frac{\xi^{2}}{r_{0}r}-1\right)  }\left(
r^{2}E-\frac{m\xi L}{\pi}\right)  ,
\end{equation}%
\begin{equation}
r\frac{d\theta}{d\tau}=\frac{1}{r\left(  \frac{\xi^{2}}{r_{0}r}-1\right)
}\left(  \frac{m\xi E}{\pi}-\left(  -1+\frac{\xi^{2}}{r_{0}r}+\left(
\frac{m\xi}{\pi r}\right)  ^{2}\right)  L\right)  .
\end{equation}
These equations of motion are derived from the principle of the least action,
$\frac{\partial S}{\partial\delta}=0$, $\frac{\partial S}{\partial E}=0$,
$\frac{\partial S}{\partial L}=0$, respectively. Under the spacetime
background of optical black holes with the metric (\ref{akm}), we consider the
massless particles which is a vortex beam emitted into the spacetime generated
by another vortex beam, and their equations of motion can be derived by taking
$\delta=0$.

\begin{figure}[ptb]
\centering
\includegraphics[width=0.8\columnwidth,height=2.5in]{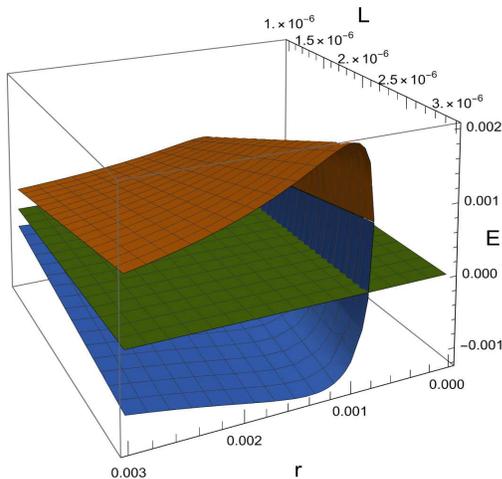} \caption{(Color
online) Energy as the function of $r$ and $L$. It indicates that the screening
of Hawking radiation with certain values of energy and angular momentum
between blue and orange surfaces. }%
\label{Fig1}%
\end{figure}

Note that $\frac{1}{2}\left(  \frac{dr}{d\tau}\right)  ^{2}$ has the meaning
of kinetic energy, so the negative effective potential makes no sense since
the velocity $\frac{dr}{d\tau}$ has to be the complex value. This means that
$\frac{dr}{d\tau}$ must be real in the calculation. However, the effective
potential (\ref{epe}) can be negative for some physically allowable values of
the energy $E$ and angular momentum $L$ under the real black-hole spacetime
background. The region in which the effective potential is negative is called
the classically forbidden region \cite{Prodanov2014,mc2016hawking}.

\begin{figure}[ptb]
\centering
\includegraphics[width=0.8\columnwidth,height=2.5in]{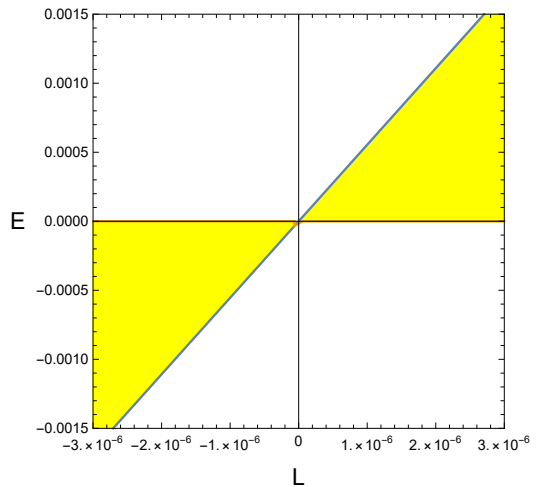} \caption{(Color
online) Energy as a function of $L$. The shielding region for particles with
energies and angular momenta at the static limit is marked with the yellow
color. }%
\label{Fig2}%
\end{figure}

Now, we analyze whether the classically forbidden region exists under the
analogue spacetime background. Fixed the angular momentum of the particle, and
thus, the effective potential can be expressed as a function of $E$,
\begin{equation}
V_{E}=r^{2}E^{2}-2\frac{m\xi EL}{\pi}+\left(  -1+\frac{\xi^{2}}{r_{0}%
r}+\left(  \frac{m\xi}{\pi r}\right)  ^{2}\right)  L^{2}. \label{eep}%
\end{equation}
It is not hard to get the negative $V_{E}$ which is derived between the two
roots due to the quadratic term of the effective potential being positive. The
classically forbidden region for values of $E$ lies between the two roots,
\begin{equation}
E_{\pm}=\frac{m\xi L}{\pi r^{2}}\left(  1\pm\sqrt{1-\frac{\frac{\xi^{2}}%
{r_{0}r}+\left(  \frac{m\xi}{\pi r}\right)  ^{2}-1}{\left(  \frac{m\xi}{\pi
r}\right)  ^{2}}}\right)  .
\end{equation}

Fig. 1 presents the classically forbidden region in which the radial motion of
the particles is restricted. In particular, the motion for the particles
within the energy range ($E_{-},E_{+}$) are constrained between the blue and
orange surfaces. This shows that the classically forbidden region exists for
the analogue black-hole spacetime. It is also noted in Fig. 1 that the
forbidden region of energy will widen when the angular momentum increases. We
also present the energy forbidden region at the static limit of the analogue
rotating black holes in Fig. 2. The results are similar to that in the Kerr
black holes \cite{mc2016hawking}, which means that the optical black hole
indeed can simulate the properties of the Kerr black holes. In what follows,
we will focus on the shielding of Penrose superradiance with an addition of
the superradiance conditions in our consideration.

\section{Penrose superradiance Shielding}

The optical black holes have the classically forbidden region, so shielding of
Penrose superradiance might cause the absence of Penrose superradiance for the
optical black holes. At first, we obtain the condition for the occurrence of
superradiance using the metric (\ref{akm}) of optical black holes.
Transforming Eq. (\ref{hje}) using the relation $\frac{d}{\mathrm{~}dr}%
=\frac{1}{r\left(  1-\frac{\xi^{2}}{r_{0}r}\right)  }\frac{d}{ds}$, we get
\begin{align}
\left(  \frac{\partial S_{r}}{\partial s}\right)  ^{2}  &  =r^{2}E^{2}%
+r^{2}\left(  \frac{\xi^{2}}{r_{0}r}-1\right)  \delta\nonumber\\
&  +\left(  -1+\frac{\xi^{2}}{r_{0}r}+\left(  \frac{m\xi}{\pi r}\right)
^{2}\right)  L^{2}-\frac{2m\xi EL}{\pi}.
\end{align}
Taking the radial wave function $\psi=\exp(iS_{r})$, the one-dimensional
Schr\"{o}dinger equation is gotten as
\begin{equation}
\frac{d^{2}\psi}{\mathrm{~}ds^{2}}+V_{eff}\psi=0.
\end{equation}
Then transform back to the coordinate system of $r$, and the similar radial
Teukolsky equation \cite{pt72,sat73} is obtained as
\begin{equation}
r\left(  1-\frac{\xi^{2}}{r_{0}r}\right)  \frac{d}{\mathrm{~}dr}\left[
r\left(  1-\frac{\xi^{2}}{r_{0}r}\right)  \frac{d\psi}{\mathrm{~}dr}\right]
+V_{eff}\psi=0.
\end{equation}

Transforming the radial coordinate $r$ into the tortoise coordinate $r^{\ast}$
by $r^{\ast}=r+\frac{\xi^{2}}{r_{0}}ln\left(  r-\frac{\xi^{2}}{r_{0}}\right)
$ and taking the new form of radial wave function $\psi^{\ast}=$ $\sqrt{r}%
\psi$, we get a linear second-order differential equation
\begin{equation}
\frac{d^{2}\psi^{\ast}}{\mathrm{~}dr^{\ast2}}+V_{eff}^{^{\prime}}\psi^{\ast
}=0,
\end{equation}
where $V_{eff}^{^{\prime}}=\frac{1}{4r^{2}}\left(  \frac{dr}{\mathrm{~}%
dr^{\ast}}\right)  ^{2}-\frac{1}{r^{2}}\left(  L^{2}+\frac{\xi^{2}}{2rr_{0}%
}\right)  \left(  \frac{dr}{\mathrm{~}dr^{\ast}}\right)  +(E-L\frac{m\xi}{\pi
r^{2}})^{2}$. It is not hard to obtain the asymptotic solution $\psi^{\ast
}=e^{iEr^{\ast}}+Re^{-iEr^{\ast}}$ for $r^{\ast}\rightarrow+\infty$,
$r\rightarrow+\infty$, $\frac{dr}{\mathrm{~}dr^{\ast}}\rightarrow1$, and
another asymptotic solution $\psi^{\ast}=Te^{i\left(  E-L\frac{m\xi}{\pi
r_{e}^{2}}\right)  r^{\ast}}$ for $r^{\ast}\rightarrow-\infty$ $r\rightarrow
r_{+}$, $\frac{dr}{\mathrm{~}dr^{\ast}}\rightarrow0$, where $R$ is the
reflection coefficient and $T$ is the transmission coefficient. Then we can
equate the Woronskian of the two asymptotic solutions to get the relation
\begin{equation}
\left\vert R\right\vert ^{2}=1-\frac{E-L\Omega_{r_{e}}}{E}\left\vert
T\right\vert ^{2},
\end{equation}
where $\Omega_{r_{e}}=\frac{m\xi}{\pi r_{e}^{2}}$ is the angular velocity at
the horizon. When $\left\vert R\right\vert ^{2}>1$, the superradiance occurs,
which leads to the superradiance condition%
\begin{equation}
E<L\Omega_{r_{e}}.\label{soc}%
\end{equation}
This is consistent with the result in Eq. (\ref{rsc}) by Zel'dovich when
taking $E=\hbar\omega$ and $L=\hbar n$ for a quantum energy and angular
momentum of the particle, respectively.

In order to incorporate the superradiance condition (\ref{soc}) in the
following discussions about the classically forbidden region. We rewrite the
effective potential (\ref{eep}) as a function of $\frac{E}{L}$,
\begin{equation}
V_{\frac{E}{L}}=r^{2}\left(  \frac{E}{L}\right)  ^{2}-2\frac{m\xi}{\pi}\left(
\frac{E}{L}\right)  +\left(  -1+\frac{\xi^{2}}{r_{0}r}+\left(  \frac{m\xi}{\pi
r}\right)  ^{2}\right)  .
\end{equation}
Using superradiance condition (\ref{soc}), we can gain the classically
forbidden region of superradiance when $V_{\frac{E}{L}}<0$. Solving the
equation $V_{\frac{E}{L}}=0$, we obtain two roots,
\begin{equation}
\left(  \frac{E}{L}\right)  _{\pm}=\frac{m\xi}{\pi r^{2}}\left(  1\pm
\sqrt{1-\frac{\frac{\xi^{2}}{r_{0}r}+\left(  \frac{m\xi}{\pi r}\right)
^{2}-1}{\left(  \frac{m\xi}{\pi r}\right)  ^{2}}}\right)  ,
\end{equation}
which gives the superradiance forbidden region between these two roots. The
upper panel of Fig. 3 presents the forbidden region under the green line
$\Omega_{r_{e}}$ but above the red line. We also mark the superradiance
forbidden region in the ergosphere with the red shadow. It is noted that there
is also the allowable region for the superradiance in the ergosphere below the
red line, and the classically forbidden region is larger than the
superradiance forbidden region between the blue line and the green line in the
ergosphere. Thus, we give the classically forbidden region inside the
ergosphere that the particles with a certain energy and angular momentum in
the red region are shielded. In particular, it requires not only the condition
(\ref{soc}) but also $n>0$ which is a supplementary condition for the
occurrence of superradiance, equivalent to $L>0$ \cite{ciszak2021}. In order
to present the influence of the angular momentum further, the effective
potential (\ref{eep}) can be reexpressed as a function of $L$,
\begin{equation}
V_{L}=\left(  -1+\frac{\xi^{2}}{r_{0}r}+\left(  \frac{m\xi}{\pi r}\right)
^{2}\right)  L^{2}-2\frac{m\xi EL}{\pi}+r^{2}E^{2}. \label{lep}%
\end{equation}
Solving the equation $V_{L}=0$, two roots are obtained,
\begin{equation}
L_{\pm}=\frac{\frac{m\xi E}{\pi}}{-1+\frac{\xi^{2}}{r_{0}r}+\left(  \frac
{m\xi}{\pi r}\right)  ^{2}}\left(  1\pm\sqrt{1-\frac{-1+\frac{\xi^{2}}{r_{0}%
r}+\left(  \frac{m\xi}{\pi r}\right)  ^{2}}{\left(  \frac{m\xi}{\pi r}\right)
^{2}}}\right)  . \label{apm}%
\end{equation}

Since the quadratic term of the equation (\ref{lep}) is negative in the
ergosphere and positive at the outside of the static limit, the classically
forbidden region of angular momentum lies between the two roots in the
ergosphere and outside the range between two roots at the outside of the
static limit. The lower panel of Fig. 3 presents the classically forbidden
region for the angular momentum. It is noted that there is also the allowable
region in the ergosphere for some specific angular momenta above the blue
line, but the forbidden region is near the static limit, which means that the
outer particles are hard to enter the ergosphere and the Hawking radiation
from the event horizon is hard to leave the ergosphere for the higher angular
momentum. Meanwhile, it is also noted that there is no shielding of the
Penrose superradiance for $L<0$ since the superradiance doesn't occur in this
region. So we get an angular momentum forbidden region between $L_{+}$ and
$L_{-}$ inside the ergosphere.

\begin{figure}[tbh]
\centering
\includegraphics[width=0.8\columnwidth,height=2.5in]{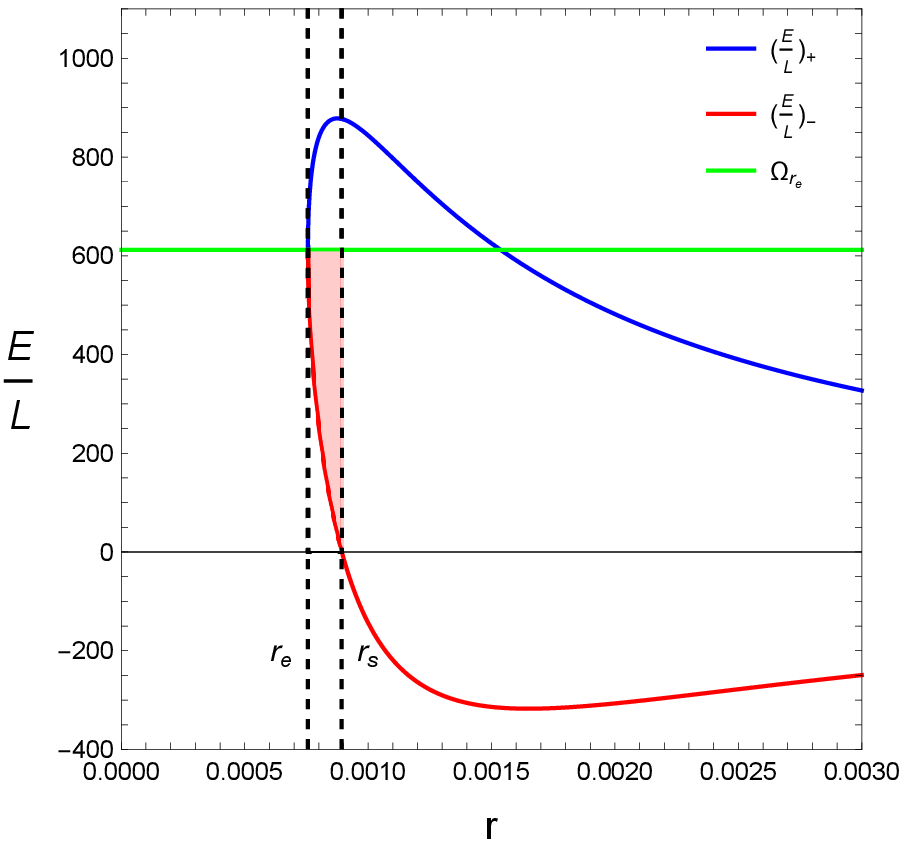}\newline%
\includegraphics[width=0.8\columnwidth,height=2.5in]{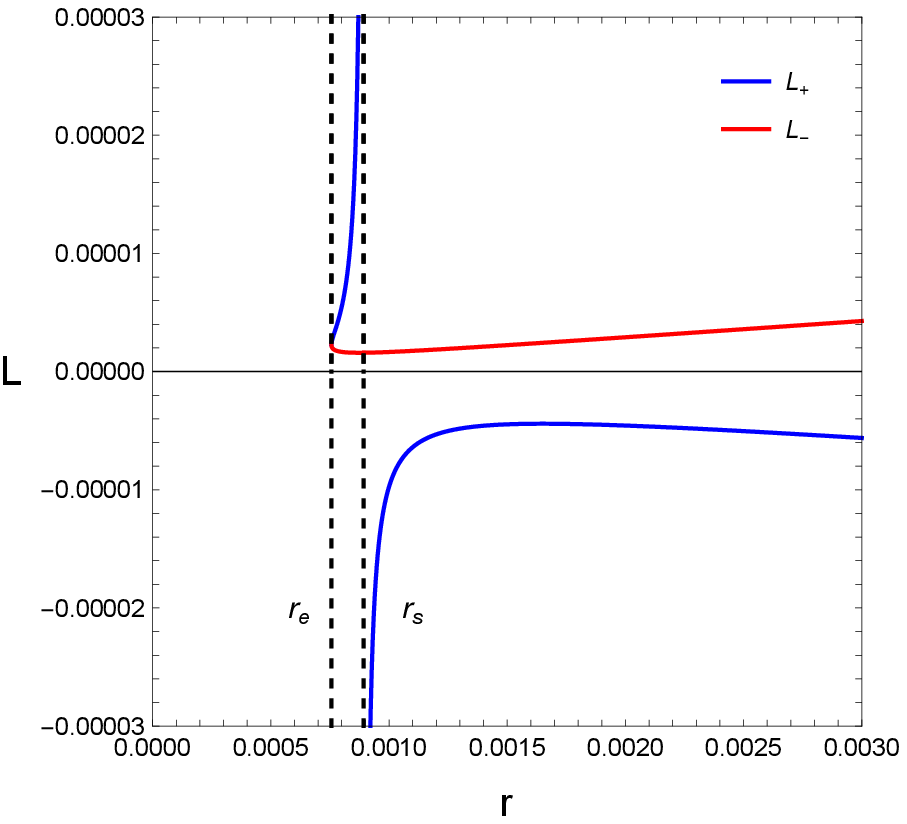}\newline%
\caption{$\frac{E}{L}$ and $L$ as the function of $r$ for the upper and lower
panel, respectively. The forbidden regions in two plots are presented with the
event horizon at $r_{e}\approx756\mu m$, the static limit at $r_{s}%
\approx893\mu m$, and the parameters $\xi=275\mu m$, $\Omega_{r_{e}}%
=\frac{v_{\theta}}{r_{e}}=612.22$. }%
\label{Fig3}%
\end{figure}

Finally, we have to point out that it is not possible to judge whether
superradiance shielding occurs in the ergosphere only based on the negative
effective potential, and it also requires to consider the superradiance
conditions $E<L\Omega_{r_{e}}$ and $L>0$. From Eq. (\ref{apm}), $L_{\pm}$ are
proportional to the energy $E$, which means that the smaller the energy, the
larger the range of forbidden region. Thus, the classical particles are harder
to realize the Penross process. Quantum tunneling is necessary for the
occurrence of Penrose superradiance with low energy. In the following section,
we will discuss this using a model related to the recent experiment for
Penrose superradiance using the optical vortices.

\section{Quantum tunneling}

As in recent numerical simulation \cite{braidotti2020} and experimental
measurement \cite{braidotti2022} about Penrose superradiance, the analogue
metric for optical black holes is written simply as%
\begin{equation}
ds^{2}\propto-(c_{s}^{2}-v_{\theta}^{2})dt^{2}+dr^{2}+(rd\theta)^{2}%
-2v_{\theta}rd\theta dt. \label{sam}%
\end{equation}
where $c_{s}$ and $v_{\theta}$ have the same forms as in the last section. In
Refs. \cite{braidotti2020,braidotti2022}, two laser beams were applied to
simulate the superradiance phenomena, in which the pump beam was used to form
the background spacetime as given in metric (\ref{sam}) and the signal beam
was transformed into an amplified output beam under the condition that an
idler beam with the negative-frequency modes trapped within the ergoregion was
generated. It is noted that the optical field of the pump beam only had an
angular phase which was used to define the angular velocity and the static
limit could be obtained by making the angular velocity be equal to the sound
speed, but no radial velocity exists which indicates the absence of the event
horizon, as presented in the metric (\ref{sam}). This is convenient to study
the phenomena derived from the existence of the static limit, which captures
the negative-energy particles and causes the Penrose superradiance.

Using the same method as in the last section, we obtain the geodesic
equations
\begin{equation}
r^{2}\left(  \frac{dr}{d\tau}\right)  ^{2}=r^{2}E^{2}-\delta r^{2}-\frac{2m\xi
EL}{\pi}+\left(  \left(  \frac{m\xi}{\pi r}\right)  ^{2}-1\right)  L^{2},
\end{equation}%
\begin{equation}
r^{2}\frac{d\theta}{d\tau}=-\frac{m\xi E}{\pi}+L\left(  \left(  \frac{m\xi
}{\pi r}\right)  ^{2}-1\right)  ,
\end{equation}%
\begin{equation}
r^{2}\frac{dt}{d\tau}=-Er^{2}+\frac{m\xi L}{\pi},
\end{equation}
and the effective potential
\begin{equation}
V_{eff}=r^{2}E^{2}-\delta r^{2}-\frac{2m\xi EL}{\pi}+\left(  \left(
\frac{m\xi}{\pi r}\right)  ^{2}-1\right)  L^{2}. \label{epw}%
\end{equation}
This effective potential can be expressed as the function of $\frac{E}{L}$,
and solving $V_{eff}\left(  \frac{E}{L}\right)  =0$, we get two roots
\begin{equation}
\left(  \frac{E}{L}\right)  _{\pm}=\frac{m\xi}{\pi r^{2}}\left(  1\pm
\sqrt{1-\frac{\left(  \frac{m\xi}{\pi r}\right)  ^{2}-1}{\left(  \frac{m\xi
}{\pi r}\right)  ^{2}}}\right)
\end{equation}
and consider the effective potential as the function of L, and solving
$V_{eff}\left(  L\right)  =0$, we get two roots
\begin{equation}
L_{\pm}=\frac{\frac{m\xi E}{\pi}}{-1+\left(  \frac{m\xi}{\pi r}\right)  ^{2}%
}\left(  1\pm\sqrt{1-\frac{-1+\left(  \frac{m\xi}{\pi r}\right)  ^{2}}{\left(
\frac{m\xi}{\pi r}\right)  ^{2}}}\right)  .
\end{equation}
By these roots, we can get the classically forbidden region for the spacetime
with the metric (\ref{sam}).

Fig. 4 presents the corresponding forbidden regions with the parameters from
Ref. \cite{braidotti2022}. For example, the linear refractive index
$n_{0}=1.32$, the nonlinearity coefficient $n_{2}=4.4\times10^{-7}cm^{2}/W$,
the wavelength $\lambda=532nm$, the pump power $P=252mW$, the value of the
pump waist $\omega_{bg}=1cm$, the pump orbital angular momentum (OAM) $l=1$,
and the signal OAM $s=2$. The pump intensity is gotten as $I=\rho_{0}%
=P/\omega_{bg}^{2}$, healing length $\xi=\lambda/\sqrt{4n_{0}\lvert
n_{2}\rvert\rho_{0}}$, the speed of sound $c_{s}=\sqrt{c^{2}\lvert n_{2}%
\rvert\rho_{0}/n_{0}^{3}}$, the flow speed $v_{\theta}=c\lvert m\rvert
/n_{0}kr$, and $m=s-l$.

\begin{figure}[tbh]
\centering
\includegraphics[width=0.8\columnwidth,height=2.5in]{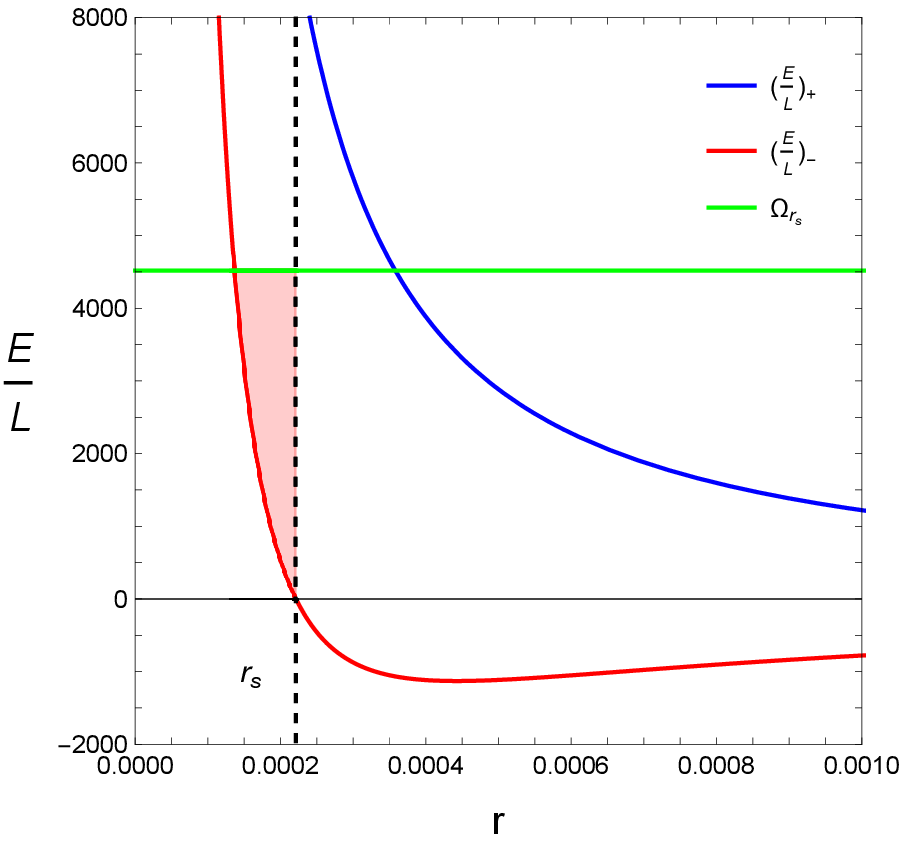}\newline%
\includegraphics[width=0.8\columnwidth,height=2.5in]{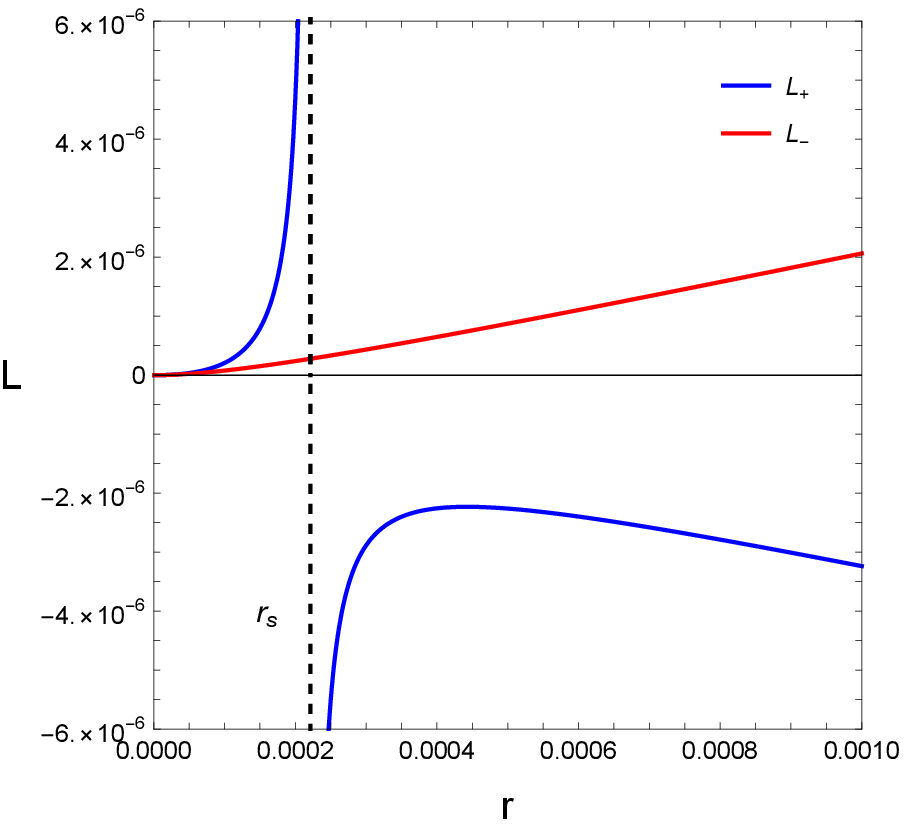}\newline%
\caption{$\frac{E}{L}$ and $L$ as the function of $r$ for the upper and lower
panel, respectively. The forbidden regions in two plots are presented with the
static limit at $r_{s}\approx221\mu m$, and the parameters $\xi=695\mu m$,
$\Omega_{r_{s}}=\frac{v_{\theta}}{r_{s}}=4518.37$. }%
\label{Fig4}%
\end{figure}

In the upper panel of Fig. 4, the superradiance forbidden region inside the
ergosphere is given with the red shadow, which is near the static limit. The
angular momentum forbidden region is presented in the lower panel of Fig. 4,
and the forbidden region inside the ergosphere is also near the static limit.
These show that the shielding of Penrose superradiance is probable although
only the static limit exists and the event horizon doesn't exist. Of course,
the superradiance shielding region has to satisfy these conditions
$E<L\Omega_{r_{s}}$ and $L>0$, as presented in the upper panel of Fig. 4.
Although there is a red forbidden region in the ergosphere, particles can
tunnel across the forbidden region.

Now consider the quantum tunneling in the analogue metric (\ref{sam}). When
$\delta=0$, the effective potential (\ref{epw}) becomes
\begin{equation}
V_{eff}=\frac{m^{2}\xi^{2}L^{2}}{\pi^{2}}\frac{1}{r^{4}}-\left(  \frac
{2mEL\xi}{\pi}+L^{2}\right)  \frac{1}{r^{2}}+E^{2}.
\end{equation}
To study the quantum tunneling, we need to know the boundary locations of the
forbidden region, which can be obtained by solving $V_{eff}=0$ for the roots
of $r$ with the fixed energy and angular momentum. Because the equation
$V_{eff}=0$ is a biquadratic equation, we can solve the roots for $x=r^{2}$,
\begin{equation}
x_{1,2}=\frac{2m\xi EL+\pi L^{2}}{2E^{2}\pi}\left(  1\pm\sqrt{1-\frac
{4E^{2}m^{2}\xi^{2}L^{2}}{\left(  2m\xi EL+\pi L^{2}\right)  ^{2}}}\right)  .
\end{equation}
x must be real, or else $r$ will take the complex values. According to
Descartes' rule of signs, when the three terms in $V_{eff}$ have the sign
$+-+$, the equation $V_{eff}=0$ has two positive roots for $x$. Thus,
\begin{equation}
r_{1}=\sqrt{x_{1}},\quad r_{2}=-\sqrt{x_{1}},\quad r_{3}=\sqrt{x_{2}},\quad
r_{4}=-\sqrt{x_{2}},
\end{equation}
are four turning points that the forbidden region changes to the allowable
region for the motion of classical particles.

The one-dimensional Schr\"{o}dinger equation with potential $V_{eff}$ is
\begin{equation}
\frac{d^{2}\psi(r)}{dr^{2}}=-V_{eff}\psi(r).
\end{equation}
The particles can tunnel through the potential barrier when $V_{eff}<0$, and
the tunneling probability is proportional to Gamow factor $e^{-2\gamma}$,
where
\begin{equation}
\gamma=\int_{r_{1}}^{r_{3}}\sqrt{-V_{eff}}dr.
\end{equation}

\begin{figure}[tbh]
\centering
\includegraphics[width=0.8\columnwidth,height=2.3in]{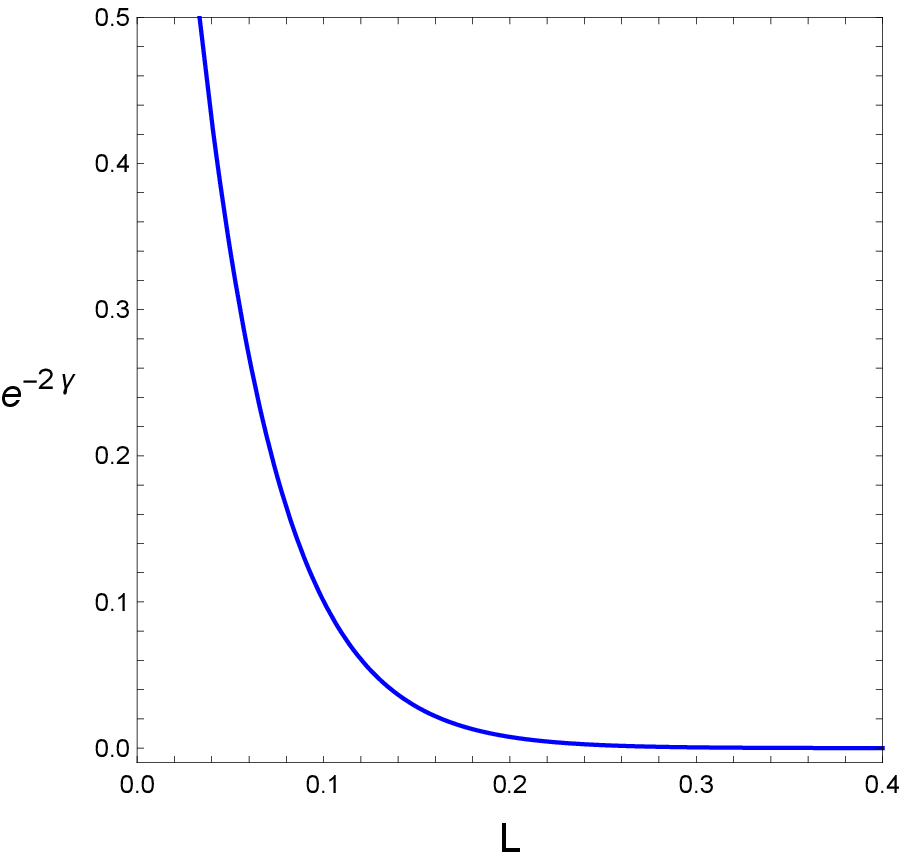}\newline%
\includegraphics[width=0.8\columnwidth,height=2.3in]{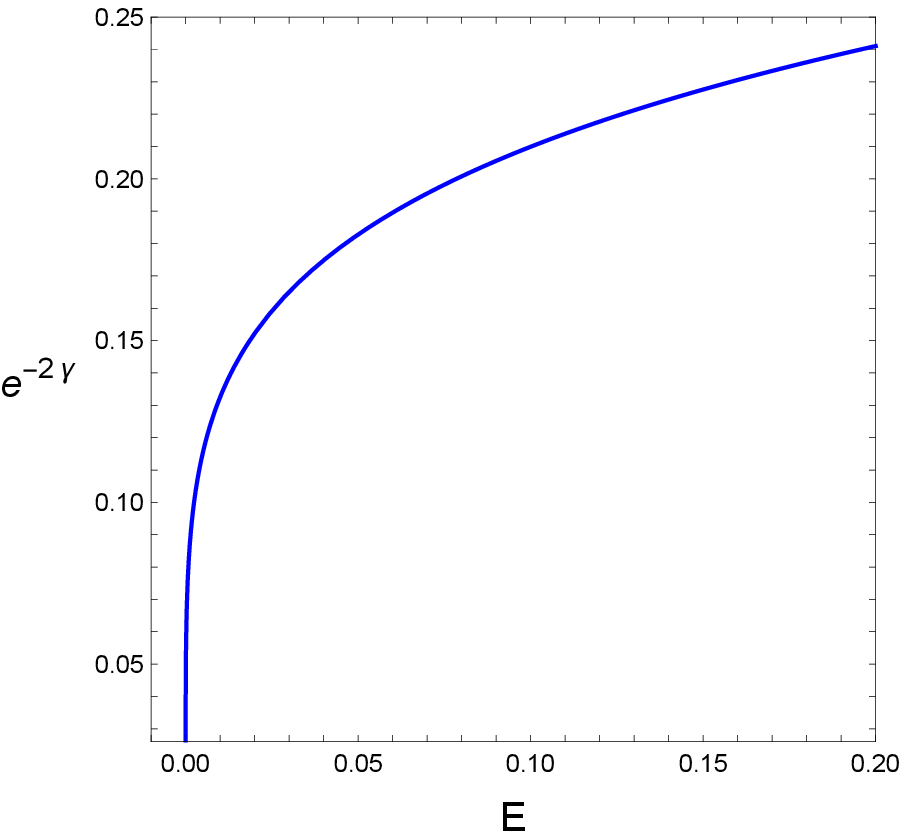}\newline%
\caption{Gamow factor as the function of angular momentum with the energy
$E=0.00252$ in the upper panel and as the function of energy with the angular
momentum $L=0.1$ in the lower panel. }%
\label{Fig5}%
\end{figure}

The distance between the turning points $r_{1}$ and $r_{3}$ is regarded as the
width of the potential barrier. The tunneling probability will decrease as the
angular momentum of the particles increases or the energy of the particles
decreases. So Penrose superradiance shielding in analogue black holes will
reduce the particles with low energies and high angular momenta, as presented
in Fig. 5.

\section{Conclusion}

In this paper, we investigate the radiation screening under the background of
optical analogue black holes and analyze mainly the Penrose superradiance
shielding for two analogue metrics, with and without the existence of the
event horizon, respectively. In order to show the existence of superradiance
shielding, we add the conditions for the occurrence of the superradiance to
analyze the classically forbidden region. We find that the forbidden region is
near the static limit for two different analogue metrics, which means that the
superradiance shielding is probable for the particles with the specific
energies and angular momenta. For the forbidden particles, we calculate the
tunneling probability, which shows that the particles with high energies and
low angular momenta are easier to tunnel through the barrier. We use the
experimental parameters to make the corresponding analyses for the forbidden
region and the tunneling, which is helpful to show that some superradiances
with the specific energies and angular momentum can be measured by the
tunneling through the static limit. Thus, these superradiances are quantum.

\section{Acknowledgment}

This work is supported from Grant No. 11654001 of the National Natural Science
Foundation of China (NSFC).

\end{document}